\documentclass[aps,prl,preprint,superscriptaddress]{revtex4-1}

\begin{document}


\title{Spectroscopy of heavy baryons with breaking of heavy-quark symmetry}


\author{S.~Yasui}
\email[]{yasuis@th.phys.titech.ac.jp}
\affiliation{Department of Physics, Tokyo Institute of Technology, Tokyo 152-8551, Japan}


\date{\today}

\begin{abstract}
Transition decay widths by one-pion emissions for excited heavy-baryons with a single heavy (charm and bottom) quark are investigated by following the heavy-quark symmetry and its breaking effects at ${\cal O}(1/M)$ for a heavy-baryon mass $M$. Based on the heavy-baryon effective theory, interaction Lagrangian for the heavy baryons with axial-vector current induced by a pion is constructed.
It is presented that the transition decay widths up to ${\cal O}(1/M)$ in several channels are constrained.
The results will be useful in experimental study of excited heavy-baryons.
\end{abstract}


\maketitle

Heavy hadrons and nuclei containing a single heavy (charm and bottom) quark give a simplified picture of QCD due to
 the heavy-quark (spin) symmetry in the heavy mass limit \cite{Isgur:1989vq,Isgur:1989ed}.
This symmetry sheds light on complicated structures in spectroscopy \cite{Isgur:1991wq}, not only for conventionally known heavy hadrons, but also for exotic heavy hadrons found recently in experiments \cite{Brambilla:2010cs} as well as hypothetical heavy-flavored nuclei \cite{Liu:2011xc,Yasui:2013vca,Yamaguchi:2013hsa,Suenaga:2014dia}.
Among them, we study excited heavy baryons ($Qqq$).
They can have relatively small decay widths even for higher spins in contrast to the light flavor sector \cite{Beringer:1900zz}.
The existence of heavy quarks clarifies motions of diquarks ($qq$) \cite{Anselmino:1992vg}, leading to color non-singlet spectroscopy \cite{Jaffe:2005md} and to a variety of diquark condensates in high density QCD matter \cite{Alford:2007xm}.
At present, however, only a few low excited states are known in experiments, and their properties are also still veiled.
We study transition decays with one-pion emissions, and find constraints imposed on them
from a view of the breaking of the heavy-quark symmetry.

Let us summarize briefly the known results, which are applicable to any heavy hadrons with a single heavy quark, in the heavy-quark mass limit \cite{Isgur:1991wq,Manohar:2000dt}.
In this limit, the heavy-quark spin $\vec{S}$ ($S=1/2$) is conserved, because the spin-dependent interactions are suppressed by inverse of the heavy-quark mass.
This is called the heavy-quark symmetry.
The total spin $\vec{J}$  of the heavy hadron containing the heavy quark is naturally conserved.
Hence, the brown muck, i.e. the light component other than the heavy quark, has a conserved total spin $\vec{j}=\vec{J}-\vec{S}$, no matter how the structure of the brown muck is complex.
We introduce the notion $\Psi^{(j)}_{J}$ for the heavy hadron with brown-muck spin $j$ and hadron total spin $J=j\pm1/2$, which is either the HQS doublet ($j\ge 1/2$) or the singlet ($j=0$). 
The two states in the HQS doublet are exactly degenerate in mass. 
The transitions are also expressed in a simple form.
Let us consider the decay with one-pion emission, $\Psi^{(j)}_{J} \!\rightarrow\! \Psi'^{(j')}_{J'}\pi$.
The decay width is given by
\begin{eqnarray}
\hspace{-1em}
\Gamma[\Psi^{(j)}_{J} \!\!\rightarrow\!\! \Psi'^{(j')}_{J'}\pi]
\propto
(2j+1)(2j'+1) \!
\left|
\left\{
\begin{array}{ccc}
 L & j'  & j  \\
 1/2  & J  & J'
\end{array}
\right\}
\right|^{2}\!\!, 
\label{eq:Isgur_Wise}
\end{eqnarray}
with $J=j\pm1/2$, $J'=j'\pm1/2$ and relative angular momentum $L$ in the final state \cite{Isgur:1991wq}.
The formula (\ref{eq:Isgur_Wise}) is applicable 
 as far as only the leading order in the $1/M$ expansion with a heavy-hadron mass $M$ is concerned.

To neglect higher orders in the expansion is, however, not always a good approximation due to the finite masses of charm and bottom quarks.
For systematic study of the higher orders,
the formalism of the heavy-hadron effective theory is useful \cite{Burdman:1992gh,Wise:1992hn,Yan:1992gz,Cho:1992gg,Casalbuoni:1996pg,Bardeen:2003kt,Manohar:2000dt}.
The applications to heavy baryons were studied \cite{Cho:1992gg,Cheng:1993gc}, and the transition decays were investigated for several excited states \cite{Pirjol:1997nh,Cheng:2006dk}.
However, they dealt with only the heavy-quark limit, whose results are naturally consistent with those obtained from Eq.~(\ref{eq:Isgur_Wise}).
Corrections at ${\cal O}(1/M)$ were investigated already in early works \cite{Cheng:1993gc}, where only the ground states with brown-muck spins zero and one were investigated.
In the present article, we study excited heavy-baryons with {\it arbitrary} brown-muck spin.

For the {\it $1/M$ corrections} for the heavy baryons, the effective Lagrangians are constructed in the following ways. 
First, we define the effective fields for heavy baryons with arbitrary brown-muck spin in the heavy-quark limit \cite{Falk:1991nq}, whose leading term in the effective Lagrangian is given by the heavy-quark symmetry.
Second, we introduce the $1/M$ corrections in the effective Lagrangian. 
We impose the invariance under the velocity rearrangement (VR), which is a boost transformation between coordinate fames with different velocities $v^{\mu}$ and $w^{\mu} = v^{\mu} + q^{\mu}/M$
with a small momentum $q^{\mu} \ll M$ \cite{Luke:1992cs,Kitazawa:1993bk}.
We also introduce the breaking terms for the heavy-quark symmetry, which is realized
 by the spin-operator acting on the heavy-quark spin.
Because the Lagrangian is uniquely determined thanks to the heavy-quark symmetry and its breaking at ${\cal O}(1/M)$, our results hold up to ${\cal O}(1/M)$ without loss of generality.
The application to the ground state was demonstrated in Ref.~\cite{Cheng:1993gc}.

To start the discussion, we introduce a tensor-spinor field for the heavy baryon with four-velocity $v^{\mu}$ and brown-muck spin $j$ \cite{Falk:1991nq}
\begin{eqnarray}
\psi^{\mu_{1}\cdots\mu_{j}} = A^{\mu_{1}\cdots\mu_{j}} u_{h},
\end{eqnarray}
defined in the heavy-quark limit,
where $u_{h}$ is a heavy-quark spinor field projected to positive-energy state
\begin{eqnarray}
v\hspace{-0.5em}/ u_{h} = u_{h},
\end{eqnarray}
and $A^{\mu_{1}\cdots\mu_{j}}$ is a tensor field of the brown muck with the following properties
\begin{eqnarray}
A^{\mu_{1}\cdots\mu_{k}\cdots\mu_{\ell}\cdots\mu_{j}} &=& A^{\mu_{1}\cdots\mu_{\ell}\cdots\mu_{k}\cdots\mu_{j}}, \\
v_{\mu_{1}} A^{\mu_{1}\cdots\mu_{j}} &=& 0, \\
g_{\mu_{1}\mu_{2}}A^{\mu_{1}\mu_{2}\cdots\mu_{j}} &=& 0.
\end{eqnarray}
Then, the conditions for $\psi^{\mu_{1}\cdots\mu_{j}}$ are given by
\begin{eqnarray}
v\hspace{-0.5em}/ \psi^{\mu_{1}\cdots\mu_{j}}  &=& \psi^{\mu_{1}\cdots\mu_{j}}, \label{eq:psi_condition_1} \\
\psi^{\mu_{1}\cdots\mu_{k}\cdots\mu_{\ell}\cdots\mu_{j}} &=& \psi^{\mu_{1}\cdots\mu_{\ell}\cdots\mu_{k}\cdots\mu_{j}}, \label{eq:psi_condition_2} \\
v_{\mu_{1}} \psi^{\mu_{1}\cdots\mu_{j}} &=& 0, \label{eq:psi_condition_3} \\
g_{\mu_{1}\mu_{2}}\psi^{\mu_{1}\mu_{2}\cdots\mu_{j}} &=& 0. \label{eq:psi_condition_4}
\end{eqnarray}
The last two conditions are required for exclusion of the irrelevant lower-spin components.
Note that $\psi^{\mu_{1}\cdots\mu_{j}}$ is a superposition of two states with total spins $j-1/2$ and $j+1/2$ in the HQS doublet ($j\neq0$) \footnote{We have only $j+1/2$ state for $j=0$.}.
The projection to each state is given as \cite{Falk:1991nq}
\begin{eqnarray}
\hspace{-2em} \psi_{j-1/2}^{\mu_{1}\cdots\mu_{j-1}} &=& \sqrt{\frac{j}{2j+1}} \gamma_{5} \gamma_{\mu_{j}} \psi^{\mu_{1}\cdots\mu_{j}}, \\
\hspace{-2em} \psi_{j+1/2}^{\mu_{1}\cdots\mu_{j}} &=& \psi^{\mu_{1}\cdots\mu_{j}} \nonumber \\ 
&-& \frac{1}{2j+1} \left( \left( \gamma^{\mu_{1}} + v^{\mu_{1}} \right) \gamma_{\nu_{1}} g^{\mu_{2}}_{\nu_{2}} \cdots g^{\mu_{j}}_{\nu_{j}} + \cdots \right. \nonumber \\
&&\left.+ g^{\mu_{1}}_{\nu_{1}}  \cdots g^{\mu_{j-1}}_{\nu_{j-1}} \left( \gamma^{\mu_{j}} + v^{\mu_{j}} \right) \gamma_{\nu_{j}} \right) \psi^{\nu_{1}\cdots\nu_{j}}. 
\end{eqnarray}
We confirm that $\psi_{j-1/2}^{\mu_{1}\cdots\mu_{j-1}}$ and $\psi_{j+1/2}^{\mu_{1}\cdots\mu_{j}}$ satisfy also
\begin{eqnarray}
\gamma_{\mu_{1}} \psi_{j-1/2}^{\mu_{1}\cdots\mu_{j-1}} &=& 0, \\
\gamma_{\mu_{1}} \psi_{j+1/2}^{\mu_{1}\cdots\mu_{j}} &=& 0,
\end{eqnarray}
in addition to the conditions (\ref{eq:psi_condition_1})-(\ref{eq:psi_condition_4}) in those $\psi^{\mu_{1}\cdots\mu_{j}}$ is replaced to $\psi_{j-1/2}^{\mu_{1}\cdots\mu_{j-1}}$ and $\psi_{j+1/2}^{\mu_{1}\cdots\mu_{j}}$.
For example, $\psi_{3/2}^{\mu_{1}}$ is the Rarita-Schwinger field with spin $3/2$ \cite{Rarita:1941mf}.

Now, we consider the interaction of heavy baryons and pions.
It is given as a coupling of tensor-spinor field $\psi^{\mu_{1}\cdots\mu_{j}}$ 
 and axial-current ${\cal A}^{\mu} = -\partial^{\mu} \hat{\pi}/f_{\pi} + {\cal O}(\hat{\pi}^{n\ge2})$ with a pion field $\hat{\pi}$ and a pion decay constant $f_{\pi}$.
In the present study, we consider the transitions between the heavy baryons with same parity.
In this case, the transitions by one-pion emissions occur by $p$-wave, and hence relatively small decay widths are expected.
For opposite parity, on the other hand,
the transitions occur by $s$-wave and the decay widths would be a sizable number, and they are hard to be observed.
We consider two cases.
One is the transitions between the different HQS doublets/singlets with brown-muck spins $j$ and $j+1$, (i) $\Psi^{(j+1)}_{J_{2}} \!\rightarrow\! \Psi^{(j)}_{J_{1}}\pi$ and (ii) $\Psi^{(j)}_{J_{1}} \!\rightarrow\! \Psi^{(j+1)}_{J_{2}}\pi$ with $J_{1}=j\pm1/2$ and $J_{2}=j+1/2, j+3/2$.
Another is the transitions between those with same brown-muck spin $j$, $\Psi^{(j)}_{J_{2}} \!\rightarrow\! \Psi'^{(j)}_{J_{1}}\pi$ with $J_{1}, J_{2}=j\pm1/2$.
We call the former the $(j,j+1)$ transitions and the latter the $(j,j)$ transitions.
We exclude the possibility of the transitions between the two states in the same HQS doublet.

First, we consider the interaction Lagrangian for the $(j,j+1)$ transitions.
Denoting the effective fields $\psi^{\mu_{1}\cdots\mu_{j}}_{1}$ and $\psi^{\mu_{1}\cdots\mu_{j+1}}_{2}$ ($j \ge 0$) for heavy baryons $\Psi^{(j)}_{J_{1}}$ and $\Psi^{(j+1)}_{J_{2}}$ \footnote{For $j=0$, we have a spinor field $\psi_{1}$ and a vector-spinor field $\psi_{2}^{\mu_{1}}$.},
the interaction Lagrangian is given by the $1/M$ expansion,
\begin{eqnarray}
{\cal L}^{(j,j+1)}_{\rm int}
&=& g^{(j,j+1)} \bar{\psi}_{1}^{\mu_{1}\cdots\mu_{j}} {\cal A}_{\mu_{j+1}} {{\psi_{2}}_{\mu_{1}\cdots\mu_{j}}}^{\mu_{j+1}} \nonumber \\
&+&
\frac{g_{1}^{(j,j+1)}}{M}
\bar{\psi}_{1}^{\mu_{1}\cdots\mu_{j}} \varepsilon_{\mu_{j+1}\rho\sigma\tau} v^{\rho} {\cal A}^{\sigma} S_{v}^{\tau} {{\psi_{2}}_{\mu_{1}\cdots\mu_{j}}}^{\mu_{j+1}} \nonumber \\
&+& {\rm h.c.} +  {\cal O}(1/M^{2}),
\label{eq:Lagrangian_jj+1}
\end{eqnarray}
with coupling constants $g^{(j,j+1)}$ and $g^{(j,j+1)}_{1}$.
The first term happens to keep the invariance under VR.
In the second term, we introduce the spin-operator (the Pauli-Lubanski vector)
$S_{v}^{\mu} = -\frac{1}{2} \gamma_{5} \left( \gamma^{\mu} v\hspace{-0.5em}/ - v^{\mu} \right)$ acting on the heavy-quark spin.
It induces the breaking of the heavy-quark symmetry.
This is understood by $S_{v_{r}}^{0}=0$ and $S_{v_{r}}^{i}=\sigma^{i}/2$ with the Pauli matrices $\sigma^{i}$ ($i=1,2,3$) in the rest frame $v_{r}=(1,\vec{0}\,)$ (cf.~Ref.~\cite{Scherer:2012xha}).

Second, we consider the interaction Lagrangian for the $(j,j)$ transitions.
Denoting the effective fields $\psi^{\mu_{1}\cdots\mu_{j}}_{1}$ and $\psi^{\mu_{1}\cdots\mu_{j}}_{2}$ for heavy baryons $\Psi'^{(j)}_{J_{1}}$ and $\Psi^{(j)}_{J_{2}}$, the interaction Lagrangian is given by the $1/M$ expansion.
For $j\ge1$, we have
\begin{eqnarray}
{\cal L}^{(j,j)}_{\rm int}
&=&
g^{(j,j)} \bar{\psi}_{1}^{\mu_{1}\cdots\mu_{j}} i \varepsilon_{\mu_{1}\rho_{1}\alpha\beta} v^{\alpha} {\cal A}^{\beta} {{\psi_{2}}^{\rho_{1}}}_{\mu_{2}\cdots\mu_{j}} \nonumber \\
&+& \frac{g^{(j,j)}}{2M} \bar{\psi}_{1}^{\mu_{1}\cdots\mu_{j}} i \varepsilon_{\mu_{1}\rho_{1}\alpha\beta} i D_{\perp}^{\alpha}(\psi_{2}) {\cal A}^{\beta} {{\psi_{2}}^{\rho_{1}}}_{\mu_{2}\cdots\mu_{j}}  \nonumber \\
&-& \frac{g^{(j,j)}}{2M} \bar{\psi}_{1}^{\mu_{1}\cdots\mu_{j}} i \varepsilon_{\mu_{1}\rho_{1}\alpha\beta} i \overleftarrow{D}_{\perp}^{\alpha}(\psi_{1}) {\cal A}^{\beta} {{\psi_{2}}^{\rho_{1}}}_{\mu_{2}\cdots\mu_{j}} \nonumber \\
&+& \frac{g_{1}^{(j,j)}}{2M} \bar{\psi}_{1}^{\mu_{1}\cdots\mu_{j}} S_{v} \!\cdot\! {\cal A} \, {\psi_{2}}_{\mu_{1}\cdots\mu_{j}} \nonumber \\
&+& \frac{g_{2}^{(j,j)}}{2M} \bar{\psi}_{1}^{\mu_{1}\cdots\mu_{j}} \left( S_{v \mu_{1}} {\cal A}_{\rho_{1}} + S_{v \rho_{1}} {\cal A}_{\mu_{1}} \right) {{\psi_{2}}^{\rho_{1}}}_{\mu_{2}\cdots\mu_{j}}  \nonumber \\
&+& {\rm h.c.} + {\cal O}(1/M^{2}),
\label{eq:Lagrangian_jj}
\end{eqnarray}
with coupling constants $g^{(j,j+1)}$, $g^{(j,j+1)}_{1}$ and $g^{(j,j+1)}_{2}$.
Here we define
$D_{\perp}^{\alpha} = D^{\alpha} - v^{\alpha} \, v \!\cdot\! D$
for a chirally covariant derivative $D^{\alpha} = \partial^{\alpha} - i{\cal V}^{\alpha}$ with a pion vector current ${\cal V}^{\alpha}$. In the above equations, $D_{\perp}^{\alpha}(\psi_{2})$ and $\overleftarrow{D}_{\perp}^{\alpha}(\psi_{1})$ indicate that they are operated to $\psi_{2}^{\rho_{1}\cdots\rho_{j}}$ and $\bar{\psi}_{1}^{\mu_{1}\cdots\mu_{j}}$, respectively.
In Eq.~(\ref{eq:Lagrangian_jj}),
the first three terms keep the invariance under VR.
The fourth and fifth terms give the breaking of the heavy-quark symmetry, because the spin-operator $S_{v}^{\mu}$ is inserted.
For $j=0$, we have
\begin{eqnarray}
{\cal L}^{(0,0)}_{\rm int} = \frac{g_{1}^{(0,0)}}{M} \bar{\psi}_{1} S_{v} \!\cdot\! {\cal A} \, \psi_{2} + {\rm h.c.} + {\cal O}(1/M^{2}),
\label{eq:Lagrangian_jj_0}
\end{eqnarray}
with a coupling constant $g^{(0,0)}_{1}$.
We note that there is no leading order term. 
Noting that the decay width becomes ${\cal O}(1/M^{2})$, we find that this is the case out of scope in the present accuracy.

Given the interaction Lagrangians (\ref{eq:Lagrangian_jj+1}) and (\ref{eq:Lagrangian_jj}), we investigate the $(j,j+1)$ and $(j,j)$ transitions, respectively.

First, we consider the $(j,j+1)$ transitions.
We have two possibilities in kinematics;
$M^{(j+1)}_{J_{2}} > M^{(j)}_{J_{1}}+m_{\pi}$ for (i),
and 
$M^{(j)}_{J_{1}} > M^{(j+1)}_{J_{2}}+m_{\pi}$ for (ii),
with $M^{(j+1)}_{J_{2}}$ ($M^{(j)}_{J_{1}}$) being a mass of $\Psi^{(j+1)}_{J_{2}}$ ($\Psi^{(j)}_{J_{1}}$) ($m_{\pi}$ a mass of a pion).
The transition decay widths $\Gamma[\Psi^{(j+1)}_{J_{2}} \!\rightarrow\! \Psi^{(j)}_{J_{1}}\pi]$ for (i) and $\Gamma[\Psi^{(j)}_{J_{1}} \!\rightarrow\! \Psi^{(j+1)}_{J_{2}}\pi]$ for (ii) are given up to ${\cal O}(1/M)$ as sums of terms with the coefficients $\left({g^{(j,j+1)}}\right)^{2}$ and $g^{(j,j+1)}g^{(j,j+1)}_{1}/M$.
Because the values of the coupling constants are not known, concrete numbers of the decay widths cannot be obtained.
However, eliminating 
 $g^{(j,j+1)}$ and $g_{1}^{(j,j+1)}$, we find the relations among the transition decay widths for each $j$ which should hold up to ${\cal O}(1/M)$.
In case (i), we obtain
\begin{eqnarray}
&& 2 \, \check{\Gamma}[\Psi^{(2)}_{3/2} \!\rightarrow\! \Psi^{(1)}_{1/2}\pi]
-4 \, \check{\Gamma}[\Psi^{(2)}_{3/2} \!\rightarrow\! \Psi^{(1)}_{3/2}\pi] \nonumber \\
&=&
\check{\Gamma}[\Psi^{(2)}_{5/2} \!\rightarrow\! \Psi^{(1)}_{3/2}\pi] 
+ {\cal O}(1/M^{2}), \label{eq:jj+1transition_i_1} \\
&& \frac{3}{2} \, \check{\Gamma}[\Psi^{(3)}_{5/2} \!\rightarrow\! \Psi^{(2)}_{3/2}\pi]
-6 \, \check{\Gamma}[\Psi^{(3)}_{5/2} \!\rightarrow\! \Psi^{(2)}_{5/2}\pi] \nonumber \\
&=&
\check{\Gamma}[\Psi^{(3)}_{7/2} \!\rightarrow\! \Psi^{(2)}_{5/2}\pi] 
+ {\cal O}(1/M^{2}), \label{eq:jj+1transition_i_2} \\
&& \frac{4}{3} \, \check{\Gamma}[\Psi^{(4)}_{7/2} \!\rightarrow\! \Psi^{(3)}_{5/2}\pi]
-8 \, \check{\Gamma}[\Psi^{(4)}_{7/2} \!\rightarrow\! \Psi^{(3)}_{7/2}] \nonumber \\
&=&
\check{\Gamma}[\Psi^{(4)}_{9/2} \!\rightarrow\! \Psi^{(3)}_{7/2}\pi] 
+ {\cal O}(1/M^{2}), \label{eq:jj+1transition_i_3}
\end{eqnarray}
where we define a dimensionless quantity
\begin{eqnarray}
\hspace{-2em} \check{\Gamma}[\Psi^{(j+1)}_{J_{2}} \!\rightarrow\! \Psi^{(j)}_{J_{1}}\pi]
= 
\frac{1}{K^{(j+1,j)}_{J_{2},J_{1}}}\, \Gamma[\Psi^{(j+1)}_{J_{2}} \!\rightarrow\! \Psi^{(j)}_{J_{1}}\pi],
\end{eqnarray}
with a kinematic factor
\begin{eqnarray}
K^{(j+1,j)}_{J_{2},J_{1}} \!=\! \frac{1}{2\pi f_{\pi}^{2}} \! \left( \left( \Delta^{(j+1,j)} \right)^{2} \!\!-\! m_{\pi}^{2} \right)^{3/2},
\label{eq:kinematic_factor}
\end{eqnarray}
with $\Delta^{(j+1,j)}={m_{\pi}^{2}}/{2M^{(j+1)}_{J_{2}}} + M^{(j+1)}_{J_{2}} - M^{(j)}_{J_{1}}$. 
The relations (\ref{eq:jj+1transition_i_1})-(\ref{eq:jj+1transition_i_3}) give constraints for transition decay widths, which hold up to ${\cal O}(1/M)$.
From the results, we may find a pattern of equations for larger $j$.
For any $j\ge1$, we find a general relation
\begin{eqnarray}
&& \frac{j+1}{j} \, \check{\Gamma}[\Psi^{(j+1)}_{j+1/2} \!\rightarrow\! \Psi^{(j)}_{j-1/2}\pi] \!-\! (2j+2) \, \check{\Gamma}[\Psi^{(j+1)}_{j+1/2} \!\rightarrow\! \Psi^{(j)}_{j+1/2}\pi] \nonumber \\
&&=
\check{\Gamma}[\Psi^{(j+1)}_{j+3/2} \!\rightarrow\! \Psi^{(j)}_{j+1/2}\pi] + {\cal O}(1/M^{2}),
\label{eq:jj+1_1_relation}
\end{eqnarray}
up to ${\cal O}(1/M)$.

In case (ii), similarly, 
 we obtain the relations among the transition decay widths as
\begin{eqnarray}
\hspace{-2em} \check{\Gamma}[\Psi^{(1)}_{1/2} \!\rightarrow\! \Psi^{(2)}_{3/2}\pi] 
&=&
4 \, \check{\Gamma}[\Psi^{(1)}_{3/2} \!\rightarrow\! \Psi^{(2)}_{3/2}\pi] \nonumber \\
&+& \frac{3}{2} \, \check{\Gamma}[\Psi^{(1)}_{3/2} \!\rightarrow\! \Psi^{(2)}_{5/2}\pi] + {\cal O}(1/M^{2}), \label{eq:jj+1transition_ii_1} \\
\hspace{-2em} \check{\Gamma}[\Psi^{(2)}_{3/2} \!\rightarrow\! \Psi^{(3)}_{5/2}\pi] 
&=&
6 \, \check{\Gamma}[\Psi^{(2)}_{5/2} \!\rightarrow\! \Psi^{(3)}_{5/2}\pi] \nonumber \\
&+& \frac{3}{4} \, \check{\Gamma}[\Psi^{(2)}_{5/2} \!\rightarrow\! \Psi^{(3)}_{7/2}\pi] + {\cal O}(1/M^{2}), \label{eq:jj+1transition_ii_2} \\
\hspace{-2em}  \check{\Gamma}[\Psi^{(3)}_{5/2} \!\rightarrow\! \Psi^{(4)}_{7/2}\pi]
&=&
8 \, \check{\Gamma}[\Psi^{(3)}_{7/2} \!\rightarrow\! \Psi^{(4)}_{7/2}\pi] \nonumber \\
&+& \frac{4}{5} \, \check{\Gamma}[\Psi^{(3)}_{7/2} \!\rightarrow\! \Psi^{(4)}_{9/2}\pi] + {\cal O}(1/M^{2}). \label{eq:jj+1transition_ii_3}
\end{eqnarray}
The relations (\ref{eq:jj+1transition_ii_1})-(\ref{eq:jj+1transition_ii_3}) again give constraints for the transition decay widths, which hold up to ${\cal O}(1/M)$.
From the results, we may expect that similar relations will hold for larger $j$.
For any $j\ge1$, we find a general relation
\begin{eqnarray}
&&\check{\Gamma}[\Psi^{(j)}_{j-1/2} \!\rightarrow\! \Psi^{(j+1)}_{j+1/2}\pi] \nonumber \\
&=&
(2j+2) \, \check{\Gamma}[\Psi^{(j)}_{j+1/2} \!\rightarrow\! \Psi^{(j+1)}_{j+1/2}\pi] \nonumber \\
&+& \frac{j+1}{j+2} \, \check{\Gamma}[\Psi^{(j)}_{j+1/2} \!\rightarrow\! \Psi^{(j+1)}_{j+3/2}\pi] 
+ {\cal O}(1/M^{2}),
\label{eq:jj+1_2_relation}
\end{eqnarray}
up to ${\cal O}(1/M)$.

Second, we consider the $(j,j)$ transitions,
supposing $M^{(j)}_{J_{2}} > M'^{(j)}_{J_{1}}+m_{\pi}$ with $M^{(j)}_{J_{2}}$ ($M'^{(j)}_{J_{1}}$) being a mass of $\Psi^{(j)}_{J_{2}}$ ($\Psi'^{(j)}_{J_{1}}$).
The analysis is given in a similar way.
For the calculated decay width $\Gamma[\Psi^{(j)}_{J_{2}} \! \rightarrow \! \Psi'^{(j)}_{J_{1}}\pi]$ for each $j$,
we eliminate 
 $g^{(j)}$, $g_{1}^{(j)}$ and $g_{2}^{(j)}$, and
use the approximation
\begin{eqnarray}
 \frac{1}{M} \left( M^{(j)}_{J_{2}} - M'^{(j)}_{J_{1}} \right)
&=&
 \frac{1}{M} \left( M^{(j)} - M'^{(j)} \right) \nonumber \\
&+& {\cal O}(1/M^2),
\end{eqnarray}
with $M^{(j)}$ ($M'^{(j)}$) being the value of $M^{(j)}_{J_{2}}$ ($M'^{(j)}_{J_{1}}$) in the heavy-quark limit.
Then, up to ${\cal O}(1/M)$, we obtain the relations for the transition decay widths as
\begin{eqnarray}
\hspace{-1.5em} \check{\Gamma}(\Psi^{(1)}_{1/2} \!\rightarrow\! \Psi'^{(1)}_{3/2}\pi)
&=&
2 \, \check{\Gamma}(\Psi^{(1)}_{3/2} \!\rightarrow\! \Psi'^{(1)}_{1/2}\pi) + {\cal O}(1/M^2), \label{eq:jjtransition_1} \\
\hspace{-1.5em} \check{\Gamma}(\Psi^{(2)}_{3/2} \!\rightarrow\! \Psi'^{(2)}_{5/2}\pi)
&=&
\frac{3}{2} \, \check{\Gamma}(\Psi^{(2)}_{5/2} \!\rightarrow\! \Psi'^{(2)}_{3/2}\pi) + {\cal O}(1/M^2), \label{eq:jjtransition_2} \\
\hspace{-1.5em} \check{\Gamma}(\Psi^{(3)}_{5/2} \!\rightarrow\! \Psi'^{(3)}_{7/2}\pi)
&=&
\frac{4}{3} \, \check{\Gamma}(\Psi^{(3)}_{7/2} \!\rightarrow\! \Psi'^{(3)}_{5/2}\pi) + {\cal O}(1/M^2), \label{eq:jjtransition_3}
\end{eqnarray}
where we define a dimensionless quantity
\begin{eqnarray}
\check{\Gamma}[\Psi^{(j)}_{J_{2}} \! \rightarrow \! \Psi'^{(j)}_{J_{1}}\pi]
=
\frac{1}{K^{(j)}_{J_{2},J_{1}}} \Gamma[\Psi^{(j)}_{J_{2}} \! \rightarrow \! \Psi'^{(j)}_{J_{1}}\pi],
\end{eqnarray}
with a kinematic factor
\begin{eqnarray}
K^{(j)}_{J_{2},J_{1}} \!=\! \frac{1}{2\pi f_{\pi}^{2}} \! \left( \left( \Delta^{(j)}_{J_{2},J_{1}} \right)^{2} \!-\! m_{\pi}^{2} \right)^{3/2},
\end{eqnarray}
with $\Delta^{(j)}_{J_{2},J_{1}}={m_{\pi}^{2}}/{2M^{(j)}_{J_{2}}}+M^{(j)}_{J_{2}} - M'^{(j)}_{J_{1}}$.
We may note the relations (\ref{eq:jjtransition_1})-(\ref{eq:jjtransition_3}) hold already at LO.
Importantly, they hold also at NLO (${\cal O}(1/M)$) without modifications.
The results suggest a pattern of relations for larger $j$.
For any $j\ge1$, we find a general relation
\begin{eqnarray}
\check{\Gamma}[\Psi^{(j)}_{j-1/2} \!\rightarrow\! \Psi'^{(j)}_{j+1/2}\pi]
&=&
\frac{j+1}{j} \, \check{\Gamma}[\Psi^{(j)}_{j+1/2} \!\rightarrow\! \Psi'^{(j)}_{j-1/2}\pi] \nonumber \\
&+& {\cal O}(1/M^2),
\label{eq:jj_relation}
\end{eqnarray}
up to ${\cal O}(1/M)$.

We have obtained the constraints among the transition decay widths, Eqs.~(\ref{eq:jj+1_1_relation}) and (\ref{eq:jj+1_2_relation}) for the $(j,j+1)$ transitions and Eq.~(\ref{eq:jj_relation}) for the $(j,j)$ transitions, which should hold up to ${\cal O}(1/M)$.
Note that those constraints are the conditions weaker than ones in the heavy-quark limit, which are given by Eq.~(\ref{eq:Isgur_Wise}).
We emphasize it important to utilize the systematic breaking of the heavy-quark symmetry, namely the invariance under VR and the heavy-quark spin operator $S_{v}^{\mu}$, as presented in the Lagrangians (\ref{eq:Lagrangian_jj+1}), (\ref{eq:Lagrangian_jj}) and (\ref{eq:Lagrangian_jj_0}).
We conclude that the results up to ${\cal O}(1/M)$ hold in a model-independent manner, because the Lagrangians are determined uniquely by construction up to this order.

We may remind us that a similar situation, that constraint relations are obtained from the breaking of symmetry, is known as the Gell-Mann-Okubo mass formula in light flavor SU(3) symmetry.
For example, we have $4m_{K}^{2} = 3m_{\eta}^{2} + m_{\pi}^{2}$ among $K$, $\eta$ and $\pi$ mesons.
This is obtained from the Gell-Mann-Oaks-Renner relation, namely the explicit breaking of the SU(3) symmetry in the meson masses,
$m_{\pi}^{2} = 2B_{0} \hat{m}$,
$m_{K}^{2} = B_{0} (\hat{m}+m_{s})$,
$m_{\eta}^{2} = \frac{2}{3} B_{0} (\hat{m}+2m_{s})$,
with a constant $B_{0}$ 
 and the averaged current mass of up and down quarks $\hat{m}$ and the current mass of strange quark $m_{s}$ (cf.~Ref.~\cite{Scherer:2012xha}). 

We consider only one-pion emission for heavy baryons.
Applications to other processes will be discussed in a similar formalism.
In any cases, the invariance under VR and the spin-operators for heavy quarks are important.

In summary, we study the transition decay widths with one-pion emissions for excited heavy-baryons with arbitrary brown-muck spin.
By considering the the $1/M$ expansion in the heavy-baryon effective theory, we find the constraints for the transition decay widths holding up to ${\cal O}(1/M)$.
Those relations are useful to explore the experimental data for the transition decay widths of excited heavy-baryons with charm and bottom quarks.
It may be worthwhile to compare our results with the quark model calculations \cite{Pirjol:1997nh,Chen:2007xf,Zhong:2007gp,Cheng:2007jx}.
However, the direct comparison is difficult, because the breaking effects of the heavy-quark symmetry were not fully considered in those references.
The detailed study will be left for future works.
Because the present approach is quite general,
similar formalisms will be applicable to heavy mesons with arbitrary brown-muck spin \cite{Falk:1991nq} and to exotic hadrons and nuclei \cite{Liu:2011xc,Yasui:2013vca,Yamaguchi:2013hsa,Suenaga:2014dia},
as far as the heavy-quark symmetry and its breaking effects are properly adopted.

\begin{acknowledgments}
We thank K.~Sudoh and M.~Oka for fruitful discussions.
This work is supported in part by Grant-in-Aid for Scientific Research on Priority Areas “Elucidation of New Hadrons with a Variety of Flavors (E01: 21105006)” and by the Grants-in-Aid for Scientific Research from JSPS (Grant No. 25247036).
\end{acknowledgments}


\begin{thebibliography}{99}
\begin{thebibliography}{32}%
\makeatletter
\providecommand \@ifxundefined [1]{%
 \@ifx{#1\undefined}
}%
\providecommand \@ifnum [1]{%
 \ifnum #1\expandafter \@firstoftwo
 \else \expandafter \@secondoftwo
 \fi
}%
\providecommand \@ifx [1]{%
 \ifx #1\expandafter \@firstoftwo
 \else \expandafter \@secondoftwo
 \fi
}%
\providecommand \natexlab [1]{#1}%
\providecommand \enquote  [1]{``#1''}%
\providecommand \bibnamefont  [1]{#1}%
\providecommand \bibfnamefont [1]{#1}%
\providecommand \citenamefont [1]{#1}%
\providecommand \href@noop [0]{\@secondoftwo}%
\providecommand \href [0]{\begingroup \@sanitize@url \@href}%
\providecommand \@href[1]{\@@startlink{#1}\@@href}%
\providecommand \@@href[1]{\endgroup#1\@@endlink}%
\providecommand \@sanitize@url [0]{\catcode `\\12\catcode `\$12\catcode
  `\&12\catcode `\#12\catcode `\^12\catcode `\_12\catcode `\%12\relax}%
\providecommand \@@startlink[1]{}%
\providecommand \@@endlink[0]{}%
\providecommand \url  [0]{\begingroup\@sanitize@url \@url }%
\providecommand \@url [1]{\endgroup\@href {#1}{\urlprefix }}%
\providecommand \urlprefix  [0]{URL }%
\providecommand \Eprint [0]{\href }%
\providecommand \doibase [0]{http://dx.doi.org/}%
\providecommand \selectlanguage [0]{\@gobble}%
\providecommand \bibinfo  [0]{\@secondoftwo}%
\providecommand \bibfield  [0]{\@secondoftwo}%
\providecommand \translation [1]{[#1]}%
\providecommand \BibitemOpen [0]{}%
\providecommand \bibitemStop [0]{}%
\providecommand \bibitemNoStop [0]{.\EOS\space}%
\providecommand \EOS [0]{\spacefactor3000\relax}%
\providecommand \BibitemShut  [1]{\csname bibitem#1\endcsname}%
\let\auto@bib@innerbib\@empty
\bibitem [{\citenamefont {Isgur}\ and\ \citenamefont
  {Wise}(1989)}]{Isgur:1989vq}%
  \BibitemOpen
  \bibfield  {author} {\bibinfo {author} {\bibfnamefont {N.}~\bibnamefont
  {Isgur}}\ and\ \bibinfo {author} {\bibfnamefont {M.~B.}\ \bibnamefont
  {Wise}},\ }\href {\doibase 10.1016/0370-2693(89)90566-2} {\bibfield
  {journal} {\bibinfo  {journal} {Phys.Lett.}\ }\textbf {\bibinfo {volume}
  {B232}},\ \bibinfo {pages} {113} (\bibinfo {year} {1989})}\BibitemShut
  {NoStop}%
\bibitem [{\citenamefont {Isgur}\ and\ \citenamefont
  {Wise}(1990)}]{Isgur:1989ed}%
  \BibitemOpen
  \bibfield  {author} {\bibinfo {author} {\bibfnamefont {N.}~\bibnamefont
  {Isgur}}\ and\ \bibinfo {author} {\bibfnamefont {M.~B.}\ \bibnamefont
  {Wise}},\ }\href {\doibase 10.1016/0370-2693(90)91219-2} {\bibfield
  {journal} {\bibinfo  {journal} {Phys.Lett.}\ }\textbf {\bibinfo {volume}
  {B237}},\ \bibinfo {pages} {527} (\bibinfo {year} {1990})}\BibitemShut
  {NoStop}%
\bibitem [{\citenamefont {Isgur}\ and\ \citenamefont
  {Wise}(1991)}]{Isgur:1991wq}%
  \BibitemOpen
  \bibfield  {author} {\bibinfo {author} {\bibfnamefont {N.}~\bibnamefont
  {Isgur}}\ and\ \bibinfo {author} {\bibfnamefont {M.~B.}\ \bibnamefont
  {Wise}},\ }\href {\doibase 10.1103/PhysRevLett.66.1130} {\bibfield  {journal}
  {\bibinfo  {journal} {Phys.Rev.Lett.}\ }\textbf {\bibinfo {volume} {66}},\
  \bibinfo {pages} {1130} (\bibinfo {year} {1991})}\BibitemShut {NoStop}%
\bibitem [{\citenamefont {Brambilla}\ \emph {et~al.}(2011)\citenamefont
  {Brambilla}, \citenamefont {Eidelman}, \citenamefont {Heltsley},
  \citenamefont {Vogt}, \citenamefont {Bodwin} \emph
  {et~al.}}]{Brambilla:2010cs}%
  \BibitemOpen
  \bibfield  {author} {\bibinfo {author} {\bibfnamefont {N.}~\bibnamefont
  {Brambilla}}, \bibinfo {author} {\bibfnamefont {S.}~\bibnamefont {Eidelman}},
  \bibinfo {author} {\bibfnamefont {B.}~\bibnamefont {Heltsley}}, \bibinfo
  {author} {\bibfnamefont {R.}~\bibnamefont {Vogt}}, \bibinfo {author}
  {\bibfnamefont {G.}~\bibnamefont {Bodwin}},  \emph {et~al.},\ }\href
  {\doibase 10.1140/epjc/s10052-010-1534-9} {\bibfield  {journal} {\bibinfo
  {journal} {Eur.Phys.J.}\ }\textbf {\bibinfo {volume} {C71}},\ \bibinfo
  {pages} {1534} (\bibinfo {year} {2011})},\ \Eprint
  {http://arxiv.org/abs/1010.5827} {arXiv:1010.5827 [hep-ph]} \BibitemShut
  {NoStop}%
\bibitem [{\citenamefont {Liu}\ and\ \citenamefont {Oka}(2012)}]{Liu:2011xc}%
  \BibitemOpen
  \bibfield  {author} {\bibinfo {author} {\bibfnamefont {Y.-R.}\ \bibnamefont
  {Liu}}\ and\ \bibinfo {author} {\bibfnamefont {M.}~\bibnamefont {Oka}},\
  }\href {\doibase 10.1103/PhysRevD.85.014015} {\bibfield  {journal} {\bibinfo
  {journal} {Phys.Rev.}\ }\textbf {\bibinfo {volume} {D85}},\ \bibinfo {pages}
  {014015} (\bibinfo {year} {2012})},\ \Eprint {http://arxiv.org/abs/1103.4624}
  {arXiv:1103.4624 [hep-ph]} \BibitemShut {NoStop}%
\bibitem [{\citenamefont {Yasui}\ \emph {et~al.}(2013)\citenamefont {Yasui},
  \citenamefont {Sudoh}, \citenamefont {Yamaguchi}, \citenamefont {Ohkoda},
  \citenamefont {Hosaka} \emph {et~al.}}]{Yasui:2013vca}%
  \BibitemOpen
  \bibfield  {author} {\bibinfo {author} {\bibfnamefont {S.}~\bibnamefont
  {Yasui}}, \bibinfo {author} {\bibfnamefont {K.}~\bibnamefont {Sudoh}},
  \bibinfo {author} {\bibfnamefont {Y.}~\bibnamefont {Yamaguchi}}, \bibinfo
  {author} {\bibfnamefont {S.}~\bibnamefont {Ohkoda}}, \bibinfo {author}
  {\bibfnamefont {A.}~\bibnamefont {Hosaka}},  \emph {et~al.},\ }\href
  {\doibase 10.1016/j.physletb.2013.10.019} {\bibfield  {journal} {\bibinfo
  {journal} {Phys.Lett.}\ }\textbf {\bibinfo {volume} {B727}},\ \bibinfo
  {pages} {185} (\bibinfo {year} {2013})},\ \Eprint
  {http://arxiv.org/abs/1304.5293} {arXiv:1304.5293 [hep-ph]} \BibitemShut
  {NoStop}%
\bibitem [{\citenamefont {Yamaguchi}\ \emph {et~al.}(2014)\citenamefont
  {Yamaguchi}, \citenamefont {Yasui},\ and\ \citenamefont
  {Hosaka}}]{Yamaguchi:2013hsa}%
  \BibitemOpen
  \bibfield  {author} {\bibinfo {author} {\bibfnamefont {Y.}~\bibnamefont
  {Yamaguchi}}, \bibinfo {author} {\bibfnamefont {S.}~\bibnamefont {Yasui}}, \
  and\ \bibinfo {author} {\bibfnamefont {A.}~\bibnamefont {Hosaka}},\ }\href
  {\doibase 10.1016/j.nuclphysa.2014.04.002} {\bibfield  {journal} {\bibinfo
  {journal} {Nucl.Phys.}\ }\textbf {\bibinfo {volume} {A927}},\ \bibinfo
  {pages} {110} (\bibinfo {year} {2014})},\ \Eprint
  {http://arxiv.org/abs/1309.4324} {arXiv:1309.4324 [nucl-th]} \BibitemShut
  {NoStop}%
\bibitem [{\citenamefont {Suenaga}\ \emph {et~al.}(2014)\citenamefont
  {Suenaga}, \citenamefont {He}, \citenamefont {Ma},\ and\ \citenamefont
  {Harada}}]{Suenaga:2014dia}%
  \BibitemOpen
  \bibfield  {author} {\bibinfo {author} {\bibfnamefont {D.}~\bibnamefont
  {Suenaga}}, \bibinfo {author} {\bibfnamefont {B.-R.}\ \bibnamefont {He}},
  \bibinfo {author} {\bibfnamefont {Y.-L.}\ \bibnamefont {Ma}}, \ and\ \bibinfo
  {author} {\bibfnamefont {M.}~\bibnamefont {Harada}},\ }\href {\doibase
  10.1103/PhysRevC.89.068201} {\bibfield  {journal} {\bibinfo  {journal}
  {Phys.Rev.}\ }\textbf {\bibinfo {volume} {C89}},\ \bibinfo {pages} {068201}
  (\bibinfo {year} {2014})},\ \Eprint {http://arxiv.org/abs/1403.5140}
  {arXiv:1403.5140 [hep-ph]} \BibitemShut {NoStop}%
\bibitem [{\citenamefont {Beringer}\ \emph {et~al.}(2012)\citenamefont
  {Beringer} \emph {et~al.}}]{Beringer:1900zz}%
  \BibitemOpen
  \bibfield  {author} {\bibinfo {author} {\bibfnamefont {J.}~\bibnamefont
  {Beringer}} \emph {et~al.} (\bibinfo {collaboration} {Particle Data Group}),\
  }\href {\doibase 10.1103/PhysRevD.86.010001} {\bibfield  {journal} {\bibinfo
  {journal} {Phys.Rev.}\ }\textbf {\bibinfo {volume} {D86}},\ \bibinfo {pages}
  {010001} (\bibinfo {year} {2012})}\BibitemShut {NoStop}%
\bibitem [{\citenamefont {Anselmino}\ \emph {et~al.}(1993)\citenamefont
  {Anselmino}, \citenamefont {Predazzi}, \citenamefont {Ekelin}, \citenamefont
  {Fredriksson},\ and\ \citenamefont {Lichtenberg}}]{Anselmino:1992vg}%
  \BibitemOpen
  \bibfield  {author} {\bibinfo {author} {\bibfnamefont {M.}~\bibnamefont
  {Anselmino}}, \bibinfo {author} {\bibfnamefont {E.}~\bibnamefont {Predazzi}},
  \bibinfo {author} {\bibfnamefont {S.}~\bibnamefont {Ekelin}}, \bibinfo
  {author} {\bibfnamefont {S.}~\bibnamefont {Fredriksson}}, \ and\ \bibinfo
  {author} {\bibfnamefont {D.}~\bibnamefont {Lichtenberg}},\ }\href {\doibase
  10.1103/RevModPhys.65.1199} {\bibfield  {journal} {\bibinfo  {journal}
  {Rev.Mod.Phys.}\ }\textbf {\bibinfo {volume} {65}},\ \bibinfo {pages} {1199}
  (\bibinfo {year} {1993})}\BibitemShut {NoStop}%
\bibitem [{\citenamefont {Jaffe}(2005)}]{Jaffe:2005md}%
  \BibitemOpen
  \bibfield  {author} {\bibinfo {author} {\bibfnamefont {R.}~\bibnamefont
  {Jaffe}},\ }\href {\doibase 10.1103/PhysRevD.72.074508} {\bibfield  {journal}
  {\bibinfo  {journal} {Phys.Rev.}\ }\textbf {\bibinfo {volume} {D72}},\
  \bibinfo {pages} {074508} (\bibinfo {year} {2005})},\ \Eprint
  {http://arxiv.org/abs/hep-ph/0507149} {arXiv:hep-ph/0507149 [hep-ph]}
  \BibitemShut {NoStop}%
\bibitem [{\citenamefont {Alford}\ \emph {et~al.}(2008)\citenamefont {Alford},
  \citenamefont {Schmitt}, \citenamefont {Rajagopal},\ and\ \citenamefont
  {Schäfer}}]{Alford:2007xm}%
  \BibitemOpen
  \bibfield  {author} {\bibinfo {author} {\bibfnamefont {M.~G.}\ \bibnamefont
  {Alford}}, \bibinfo {author} {\bibfnamefont {A.}~\bibnamefont {Schmitt}},
  \bibinfo {author} {\bibfnamefont {K.}~\bibnamefont {Rajagopal}}, \ and\
  \bibinfo {author} {\bibfnamefont {T.}~\bibnamefont {Schäfer}},\ }\href
  {\doibase 10.1103/RevModPhys.80.1455} {\bibfield  {journal} {\bibinfo
  {journal} {Rev.Mod.Phys.}\ }\textbf {\bibinfo {volume} {80}},\ \bibinfo
  {pages} {1455} (\bibinfo {year} {2008})},\ \Eprint
  {http://arxiv.org/abs/0709.4635} {arXiv:0709.4635 [hep-ph]} \BibitemShut
  {NoStop}%
\bibitem [{\citenamefont {Manohar}\ and\ \citenamefont
  {Wise}(2000)}]{Manohar:2000dt}%
  \BibitemOpen
  \bibfield  {author} {\bibinfo {author} {\bibfnamefont {A.~V.}\ \bibnamefont
  {Manohar}}\ and\ \bibinfo {author} {\bibfnamefont {M.~B.}\ \bibnamefont
  {Wise}},\ }\href@noop {} {\bibfield  {journal} {\bibinfo  {journal}
  {Camb.Monogr.Part.Phys.Nucl.Phys.Cosmol.}\ }\textbf {\bibinfo {volume}
  {10}},\ \bibinfo {pages} {1} (\bibinfo {year} {2000})}\BibitemShut {NoStop}%
\bibitem [{\citenamefont {Burdman}\ and\ \citenamefont
  {Donoghue}(1992)}]{Burdman:1992gh}%
  \BibitemOpen
  \bibfield  {author} {\bibinfo {author} {\bibfnamefont {G.}~\bibnamefont
  {Burdman}}\ and\ \bibinfo {author} {\bibfnamefont {J.~F.}\ \bibnamefont
  {Donoghue}},\ }\href {\doibase 10.1016/0370-2693(92)90068-F} {\bibfield
  {journal} {\bibinfo  {journal} {Phys.Lett.}\ }\textbf {\bibinfo {volume}
  {B280}},\ \bibinfo {pages} {287} (\bibinfo {year} {1992})}\BibitemShut
  {NoStop}%
\bibitem [{\citenamefont {Wise}(1992)}]{Wise:1992hn}%
  \BibitemOpen
  \bibfield  {author} {\bibinfo {author} {\bibfnamefont {M.~B.}\ \bibnamefont
  {Wise}},\ }\href {\doibase 10.1103/PhysRevD.45.R2188} {\bibfield  {journal}
  {\bibinfo  {journal} {Phys.Rev.}\ }\textbf {\bibinfo {volume} {D45}},\
  \bibinfo {pages} {2188} (\bibinfo {year} {1992})}\BibitemShut {NoStop}%
\bibitem [{\citenamefont {Yan}\ \emph {et~al.}(1992)\citenamefont {Yan},
  \citenamefont {Cheng}, \citenamefont {Cheung}, \citenamefont {Lin},
  \citenamefont {Lin} \emph {et~al.}}]{Yan:1992gz}%
  \BibitemOpen
  \bibfield  {author} {\bibinfo {author} {\bibfnamefont {T.-M.}\ \bibnamefont
  {Yan}}, \bibinfo {author} {\bibfnamefont {H.-Y.}\ \bibnamefont {Cheng}},
  \bibinfo {author} {\bibfnamefont {C.-Y.}\ \bibnamefont {Cheung}}, \bibinfo
  {author} {\bibfnamefont {G.-L.}\ \bibnamefont {Lin}}, \bibinfo {author}
  {\bibfnamefont {Y.}~\bibnamefont {Lin}},  \emph {et~al.},\ }\href {\doibase
  10.1103/PhysRevD.46.1148, 10.1103/PhysRevD.55.5851} {\bibfield  {journal}
  {\bibinfo  {journal} {Phys.Rev.}\ }\textbf {\bibinfo {volume} {D46}},\
  \bibinfo {pages} {1148} (\bibinfo {year} {1992})}\BibitemShut {NoStop}%
\bibitem [{\citenamefont {Cho}(1992)}]{Cho:1992gg}%
  \BibitemOpen
  \bibfield  {author} {\bibinfo {author} {\bibfnamefont {P.~L.}\ \bibnamefont
  {Cho}},\ }\href {\doibase 10.1016/0370-2693(92)91314-Y} {\bibfield  {journal}
  {\bibinfo  {journal} {Phys.Lett.}\ }\textbf {\bibinfo {volume} {B285}},\
  \bibinfo {pages} {145} (\bibinfo {year} {1992})},\ \Eprint
  {http://arxiv.org/abs/hep-ph/9203225} {arXiv:hep-ph/9203225 [hep-ph]}
  \BibitemShut {NoStop}%
\bibitem [{\citenamefont {Casalbuoni}\ \emph {et~al.}(1997)\citenamefont
  {Casalbuoni}, \citenamefont {Deandrea}, \citenamefont {Di~Bartolomeo},
  \citenamefont {Gatto}, \citenamefont {Feruglio} \emph
  {et~al.}}]{Casalbuoni:1996pg}%
  \BibitemOpen
  \bibfield  {author} {\bibinfo {author} {\bibfnamefont {R.}~\bibnamefont
  {Casalbuoni}}, \bibinfo {author} {\bibfnamefont {A.}~\bibnamefont
  {Deandrea}}, \bibinfo {author} {\bibfnamefont {N.}~\bibnamefont
  {Di~Bartolomeo}}, \bibinfo {author} {\bibfnamefont {R.}~\bibnamefont
  {Gatto}}, \bibinfo {author} {\bibfnamefont {F.}~\bibnamefont {Feruglio}},
  \emph {et~al.},\ }\href {\doibase 10.1016/S0370-1573(96)00027-0} {\bibfield
  {journal} {\bibinfo  {journal} {Phys.Rept.}\ }\textbf {\bibinfo {volume}
  {281}},\ \bibinfo {pages} {145} (\bibinfo {year} {1997})},\ \Eprint
  {http://arxiv.org/abs/hep-ph/9605342} {arXiv:hep-ph/9605342 [hep-ph]}
  \BibitemShut {NoStop}%
\bibitem [{\citenamefont {Bardeen}\ \emph {et~al.}(2003)\citenamefont
  {Bardeen}, \citenamefont {Eichten},\ and\ \citenamefont
  {Hill}}]{Bardeen:2003kt}%
  \BibitemOpen
  \bibfield  {author} {\bibinfo {author} {\bibfnamefont {W.~A.}\ \bibnamefont
  {Bardeen}}, \bibinfo {author} {\bibfnamefont {E.~J.}\ \bibnamefont
  {Eichten}}, \ and\ \bibinfo {author} {\bibfnamefont {C.~T.}\ \bibnamefont
  {Hill}},\ }\href {\doibase 10.1103/PhysRevD.68.054024} {\bibfield  {journal}
  {\bibinfo  {journal} {Phys.Rev.}\ }\textbf {\bibinfo {volume} {D68}},\
  \bibinfo {pages} {054024} (\bibinfo {year} {2003})},\ \Eprint
  {http://arxiv.org/abs/hep-ph/0305049} {arXiv:hep-ph/0305049 [hep-ph]}
  \BibitemShut {NoStop}%
\bibitem [{\citenamefont {Cheng}\ \emph {et~al.}(1994)\citenamefont {Cheng},
  \citenamefont {Cheung}, \citenamefont {Lin}, \citenamefont {Lin},
  \citenamefont {Yan} \emph {et~al.}}]{Cheng:1993gc}%
  \BibitemOpen
  \bibfield  {author} {\bibinfo {author} {\bibfnamefont {H.-Y.}\ \bibnamefont
  {Cheng}}, \bibinfo {author} {\bibfnamefont {C.-Y.}\ \bibnamefont {Cheung}},
  \bibinfo {author} {\bibfnamefont {G.-L.}\ \bibnamefont {Lin}}, \bibinfo
  {author} {\bibfnamefont {Y.}~\bibnamefont {Lin}}, \bibinfo {author}
  {\bibfnamefont {T.-M.}\ \bibnamefont {Yan}},  \emph {et~al.},\ }\href
  {\doibase 10.1103/PhysRevD.49.2490} {\bibfield  {journal} {\bibinfo
  {journal} {Phys.Rev.}\ }\textbf {\bibinfo {volume} {D49}},\ \bibinfo {pages}
  {2490} (\bibinfo {year} {1994})},\ \Eprint
  {http://arxiv.org/abs/hep-ph/9308283} {arXiv:hep-ph/9308283 [hep-ph]}
  \BibitemShut {NoStop}%
\bibitem [{\citenamefont {Pirjol}\ and\ \citenamefont
  {Yan}(1997)}]{Pirjol:1997nh}%
  \BibitemOpen
  \bibfield  {author} {\bibinfo {author} {\bibfnamefont {D.}~\bibnamefont
  {Pirjol}}\ and\ \bibinfo {author} {\bibfnamefont {T.-M.}\ \bibnamefont
  {Yan}},\ }\href {\doibase 10.1103/PhysRevD.56.5483} {\bibfield  {journal}
  {\bibinfo  {journal} {Phys.Rev.}\ }\textbf {\bibinfo {volume} {D56}},\
  \bibinfo {pages} {5483} (\bibinfo {year} {1997})},\ \Eprint
  {http://arxiv.org/abs/hep-ph/9701291} {arXiv:hep-ph/9701291 [hep-ph]}
  \BibitemShut {NoStop}%
\bibitem [{\citenamefont {Cheng}\ and\ \citenamefont
  {Chua}(2007)}]{Cheng:2006dk}%
  \BibitemOpen
  \bibfield  {author} {\bibinfo {author} {\bibfnamefont {H.-Y.}\ \bibnamefont
  {Cheng}}\ and\ \bibinfo {author} {\bibfnamefont {C.-K.}\ \bibnamefont
  {Chua}},\ }\href {\doibase 10.1103/PhysRevD.75.014006} {\bibfield  {journal}
  {\bibinfo  {journal} {Phys.Rev.}\ }\textbf {\bibinfo {volume} {D75}},\
  \bibinfo {pages} {014006} (\bibinfo {year} {2007})},\ \Eprint
  {http://arxiv.org/abs/hep-ph/0610283} {arXiv:hep-ph/0610283 [hep-ph]}
  \BibitemShut {NoStop}%
\bibitem [{\citenamefont {Falk}(1992)}]{Falk:1991nq}%
  \BibitemOpen
  \bibfield  {author} {\bibinfo {author} {\bibfnamefont {A.~F.}\ \bibnamefont
  {Falk}},\ }\href {\doibase 10.1016/0550-3213(92)90004-U} {\bibfield
  {journal} {\bibinfo  {journal} {Nucl.Phys.}\ }\textbf {\bibinfo {volume}
  {B378}},\ \bibinfo {pages} {79} (\bibinfo {year} {1992})}\BibitemShut
  {NoStop}%
\bibitem [{\citenamefont {Luke}\ and\ \citenamefont
  {Manohar}(1992)}]{Luke:1992cs}%
  \BibitemOpen
  \bibfield  {author} {\bibinfo {author} {\bibfnamefont {M.~E.}\ \bibnamefont
  {Luke}}\ and\ \bibinfo {author} {\bibfnamefont {A.~V.}\ \bibnamefont
  {Manohar}},\ }\href {\doibase 10.1016/0370-2693(92)91786-9} {\bibfield
  {journal} {\bibinfo  {journal} {Phys.Lett.}\ }\textbf {\bibinfo {volume}
  {B286}},\ \bibinfo {pages} {348} (\bibinfo {year} {1992})},\ \Eprint
  {http://arxiv.org/abs/hep-ph/9205228} {arXiv:hep-ph/9205228 [hep-ph]}
  \BibitemShut {NoStop}%
\bibitem [{\citenamefont {Kitazawa}\ and\ \citenamefont
  {Kurimoto}(1994)}]{Kitazawa:1993bk}%
  \BibitemOpen
  \bibfield  {author} {\bibinfo {author} {\bibfnamefont {N.}~\bibnamefont
  {Kitazawa}}\ and\ \bibinfo {author} {\bibfnamefont {T.}~\bibnamefont
  {Kurimoto}},\ }\href {\doibase 10.1016/0370-2693(94)00047-6} {\bibfield
  {journal} {\bibinfo  {journal} {Phys.Lett.}\ }\textbf {\bibinfo {volume}
  {B323}},\ \bibinfo {pages} {65} (\bibinfo {year} {1994})},\ \Eprint
  {http://arxiv.org/abs/hep-ph/9312225} {arXiv:hep-ph/9312225 [hep-ph]}
  \BibitemShut {NoStop}%
\bibitem [{Note1()}]{Note1}%
  \BibitemOpen
  \bibinfo {note} {We have only $j+1/2$ state for $j=0$.}\BibitemShut {Stop}%
\bibitem [{\citenamefont {Rarita}\ and\ \citenamefont
  {Schwinger}(1941)}]{Rarita:1941mf}%
  \BibitemOpen
  \bibfield  {author} {\bibinfo {author} {\bibfnamefont {W.}~\bibnamefont
  {Rarita}}\ and\ \bibinfo {author} {\bibfnamefont {J.}~\bibnamefont
  {Schwinger}},\ }\href {\doibase 10.1103/PhysRev.60.61} {\bibfield  {journal}
  {\bibinfo  {journal} {Phys.Rev.}\ }\textbf {\bibinfo {volume} {60}},\
  \bibinfo {pages} {61} (\bibinfo {year} {1941})}\BibitemShut {NoStop}%
\bibitem [{Note2()}]{Note2}%
  \BibitemOpen
  \bibinfo {note} {For $j=0$, we have a spinor field $\psi _{1}$ and a
  vector-spinor field $\psi _{2}^{\mu _{1}}$.}\BibitemShut {Stop}%
\bibitem [{\citenamefont {Scherer}\ and\ \citenamefont
  {Schindler}(2012)}]{Scherer:2012xha}%
  \BibitemOpen
  \bibfield  {author} {\bibinfo {author} {\bibfnamefont {S.}~\bibnamefont
  {Scherer}}\ and\ \bibinfo {author} {\bibfnamefont {M.~R.}\ \bibnamefont
  {Schindler}},\ }\href {\doibase 10.1007/978-3-642-19254-8} {\bibfield
  {journal} {\bibinfo  {journal} {Lect.Notes Phys.}\ }\textbf {\bibinfo
  {volume} {830}},\ \bibinfo {pages} {pp.1} (\bibinfo {year}
  {2012})}\BibitemShut {NoStop}%
\bibitem [{\citenamefont {Chen}\ \emph {et~al.}(2007)\citenamefont {Chen},
  \citenamefont {Chen}, \citenamefont {Liu}, \citenamefont {Deng},\ and\
  \citenamefont {Zhu}}]{Chen:2007xf}%
  \BibitemOpen
  \bibfield  {author} {\bibinfo {author} {\bibfnamefont {C.}~\bibnamefont
  {Chen}}, \bibinfo {author} {\bibfnamefont {X.-L.}\ \bibnamefont {Chen}},
  \bibinfo {author} {\bibfnamefont {X.}~\bibnamefont {Liu}}, \bibinfo {author}
  {\bibfnamefont {W.-Z.}\ \bibnamefont {Deng}}, \ and\ \bibinfo {author}
  {\bibfnamefont {S.-L.}\ \bibnamefont {Zhu}},\ }\href {\doibase
  10.1103/PhysRevD.75.094017} {\bibfield  {journal} {\bibinfo  {journal}
  {Phys.Rev.}\ }\textbf {\bibinfo {volume} {D75}},\ \bibinfo {pages} {094017}
  (\bibinfo {year} {2007})},\ \Eprint {http://arxiv.org/abs/0704.0075}
  {arXiv:0704.0075 [hep-ph]} \BibitemShut {NoStop}%
\bibitem [{\citenamefont {Zhong}\ and\ \citenamefont
  {Zhao}(2008)}]{Zhong:2007gp}%
  \BibitemOpen
  \bibfield  {author} {\bibinfo {author} {\bibfnamefont {X.-H.}\ \bibnamefont
  {Zhong}}\ and\ \bibinfo {author} {\bibfnamefont {Q.}~\bibnamefont {Zhao}},\
  }\href {\doibase 10.1103/PhysRevD.77.074008} {\bibfield  {journal} {\bibinfo
  {journal} {Phys.Rev.}\ }\textbf {\bibinfo {volume} {D77}},\ \bibinfo {pages}
  {074008} (\bibinfo {year} {2008})},\ \Eprint {http://arxiv.org/abs/0711.4645}
  {arXiv:0711.4645 [hep-ph]} \BibitemShut {NoStop}%
\bibitem [{\citenamefont {Cheng}(2007)}]{Cheng:2007jx}%
  \BibitemOpen
  \bibfield  {author} {\bibinfo {author} {\bibfnamefont {H.-Y.}\ \bibnamefont
  {Cheng}},\ }\href@noop {} {\bibfield  {journal} {\bibinfo  {journal} {eConf}\
  }\textbf {\bibinfo {volume} {C070805}},\ \bibinfo {pages} {35} (\bibinfo
  {year} {2007})},\ \Eprint {http://arxiv.org/abs/0709.0958} {arXiv:0709.0958
  [hep-ph]} \BibitemShut {NoStop}%
\end{thebibliography}
%

\end{thebibliography}

\end{document}